\documentclass[12pt]{article}
\usepackage{graphics}
\usepackage{amsthm}
\usepackage{amssymb}
\usepackage{amsmath}
\usepackage{amsfonts}
\usepackage{epic}
\usepackage{hyperref}
\usepackage{epsfig}
\usepackage{bm}
\usepackage{amscd}

\theoremstyle{plain}

\theoremstyle{definition}

\def\be{\begin{equation}}
\def\ee{\end{equation}}
\smallskip

\begin{document}

\headsep=-0.5cm

\begin{titlepage}
\begin{flushright}
%hep-th/.......\\
\end{flushright}
%%%%%%%%%%%%%%%%%%%%%%%%%%%%%%%%%%%%%%%%%%%%%%%%%%%%%%%%%%%%%%%%%%%%%%%%
\begin{center}
\noindent{{\LARGE{Zooming in on the horizon when in its Meissner state}}}

\smallskip
\smallskip
\smallskip

\smallskip
\smallskip

\smallskip

\smallskip

\smallskip
\smallskip
\noindent{\large{Gaston Giribet$^{1}$, Joan La Madrid$^{1}$, Luciano Montecchio$^{2}$, Emilio Rub\'{\i}n de Celis$^{2}$, Pedro Schmied$^{2}$}}
\end{center}

\smallskip

\centerline{$^1$ Department of Physics, New York University, NYU}
\centerline{{\it 726 Broadway, New York, NY10003, USA.}}

\smallskip
\smallskip

\centerline{$^2$ Department of Physics, University of Buenos Aires, UBA and IFIBA, CONICET}
\centerline{{\it Ciudad Universitaria, pabell\'on 1, 1428, Buenos Aires, Argentina.}}

\smallskip
\smallskip

\smallskip
\smallskip

\smallskip

\smallskip

\begin{abstract}
When approaching extremality, rotating black holes tend to expel the magnetic field in which they are immersed. This phenomenon, being reminiscent of the Meissner-Ochsenfeld effect in superconductors, is known as the black hole Meissner effect, and here we study it in the backreacting regime and from the near horizon perspective. By resorting to methods recently developed in the literature, which allow to compute conserved charges in the near horizon region, regardless the details of the asymptotia at large distance, we investigate the properties of the black hole horizon when in its Meissner state. We show that, when in such state, the horizon exhibits two sets of supertranslation symmetries as well as a symmetry generated by the local conformal group. The supertranslations are generated by two infinite sets of currents, one of which comes from local dilations of the advanced null coordinate at the horizon, and the other from local gauge transformations that preserve the electromagnetic field configuration at the horizon. We show that the evaluation of the conserved charges associated to these symmetries correctly reproduce the physical charges of the magnetized black holes and their thermodynamics. This represents a concrete application of the techniques developed in \cite{Laura, Extended, Puhm} and it extends the results of \cite{Lucho} to arbitrary values of the black hole charges. In addition, we elaborate on the charges computation at the horizon: we show the equivalence between the horizon charges and the evaluation of the corresponding Komar integrals. Besides, we show the validity of the Gauss phenomenon by explicitly relating near horizon charges with fluxes and charges computed by other techniques. All this provides a method to derive the thermodynamics of magnetized horizons in a quite succinct way, including the case of horizons exhibiting the Meissner effect.  

\end{abstract}

\end{titlepage}
%%%%%%%%%%%%%%%%%%%%%%%%%%%%%%%%%%%%%%%%%%%%%%%%%%%%%%%%%%%%%%%%%%%%%%%%

\newpage

%\tableofcontents

%\newpage

%%%%%%%%%%%%%%%%%%%%%%%%%%%%%%%%%%%%%%%%%%%%%%%%%%%%%%%%%%%%%%%%%%%%%%

%\maketitle

\section{Introduction}

Black hole event horizons are one of the most intriguing objects in physics. Thinking about their existence frequently leads us to reconsider what we think we understand about basal concepts such as causality, locality, predictability, information, and unitarity. In recent years, the nature of event horizons has been intensively studied in both mathematical physics and astrophysics, and this has led us to interesting speculations about the structure of spacetime in their vicinity: the puzzle of information loss \cite{information}, the firewall paradox \cite{firewall}, the black hole complementarity \cite{complementarity}, the proposal of fuzzballs \cite{fuzzball}, and the discovery of conformal symmetries in the near horizon geometry \cite{KerrCFT} are among the most interesting subjects that have attracted attention in high energy physics research in the last decades. In the astrophysical context, on the other hand, the new possibility of having observational access at the scale of the event horizon initiated a new era: the quantitative study of the black hole shadow \cite{shadow}, the analysis of the inner disk dynamics \cite{dynamics}, the observation of the photon ring \cite{ring} and of the polarization produced by the magnetic field near the horizon \cite{EHT} allow relativistic astrophysics to be tested in a regime hitherto unimagined. All this motivates the detailed study of the physical processes that take place in the vicinity of black hole event horizons. 

Among all the interesting phenomena that take place near the event horizons is the Meissner effect \cite{King}; that is, the fact that, under certain conditions, black holes in magnetic fields behave as superconductors do. Near extremality, when their Hawking temperature goes to zero, spinning black holes tend to eject the lines of magnetic fields from their event horizons. This is analogous to the Meissner-Ochsenfeld effect in superconductors \cite{Meissner}; namely, the expulsion of the magnetic field from a superconducting material during its transition to the superconducting state. It has been argued that studying the Meissner effect in black holes might be of importance in the astrophysical context \cite{Penna, Mac, Mac1, Mac2, Mac3, Mac4}, especially if it is considered in connection to the so-called Blandford-Znajek process \cite{BZ}, that is, the process by which energy can be extracted from the rotation of a black hole and transferred to the generation of relativistic jets. In the Blandford-Znajek mechanism, the ergosphere plays an important role, causing the magnetosphere within it to rotate, ultimately resulting in the extraction of angular momentum from the black hole. Since the Blandford-Znajek mechanism is the favoured explanation for the jet generation in quasars and other sources, concern arises as whether the Meissner effect could quench the power of the jets in the case of rapidly rotating black holes, as the magnetic field is necessary for the entire process to develop. This has recently been discussed in the literature \cite{Penna}, where it has been argued that the feeding process can actually continue all the way to the extremal limit, and therefore the jets are not necessarily turned off by the Meissner effect. This seems to be in agreement with relativistic magnetohydrodynamics simulations as well as with observations of near-extremal black hole candidates. 

While lately there have been important advances in the near horizon magnetohydrodynamics, further analytic studies of the role played by the magnetic field in the zone close to the spinning black holes are necessary. Here, we will study the black hole Meissner effect from the near horizon perspective and in the regime in which the magnetic field is fully backreacting on the spacetime geometry. By resorting to methods recently developed in the literature \cite{Laura, Extended}, which allow to compute conserved charges in the near horizon region, we will investigate the properties of the black hole horizon when in its Meissner state. We will show that, when in such state, the horizon exhibits infinite-dimensional symmetries: two sets of supertranslation symmetries as well as a symmetry generated by the local conformal group. The supertranslations are generated by two infinite sets of currents, one of which comes from local dilations of the advanced null coordinate $v$ at the horizon $H$, and the other from local gauge transformations that preserve the electromagnetic field configuration at the horizon. As we will show, the evaluation of the Noether charges associated to the zero modes of these symmetries correctly reproduces the black hole physical charges and its thermodynamics. This represents a concrete application of the techniques developed in \cite{Puhm} and it extends the results of \cite{Lucho} to the case of arbitrary values of the black hole charges. In addition, we will elaborate on the charges computation at the horizon: we will show that the horizon charges admit to be written as Komar integrals on $H$. Besides, we will explicitly show that the near horizon charges can be written as flux integrals, explaining in this way the agreement with the computations performed with more traditional methods, some of which require to handle the asymptotic conditions at large distance. This analysis will enable us to derive the thermodynamics of black holes in magnetic environments in a remarkably succinct manner, and then we will apply this to the case of black holes exhibiting Meissner effect. 

Our paper is organized as follows: In section 2, we study the near horizon geometry of the Kerr-Newman black hole immersed in a backreacting magnetic field. This is given by a limit of the Ernst-Wild solution to Einstein-Maxwell equation, which describes an electrically charged spinning black hole embedded in a Melvin universe. We study the symmetries of the solution and show that it admits an infinite-dimensional set of asymptotic Killing vectors that preserve the near horizon boundary conditions. In section 3, we analyze the Noether charges associated to the near horizon symmetries of the magnetized black holes. We show that these {\it horizon charges} can be expressed as Komar integrals and admit to be written as flux integrals. The latter proves the validity of the Gauss law and explains the success of the near horizon method. In section 4, we apply the study of the Noether charges and thermodynamics to the case of event horizons of black holes when in their Meissner states. We derive their thermodynamics and we discuss the emergence of an infinite-dimensional symmetry in their vicinity.

The black hole Meissner effect was also studied recently in references \cite{me1, me2, me3, me4, me5, me6, me7, me8, me9, me11, Ruffini, Bicak, Bicak2, Mac, Voro, Budin, backreaction}; see also references thereof; from the near horizon perspective, it was studied in \cite{AstorinoKerrCFT, otroKerrCFT, ne1, ne2, ne3, ne4}, although with approaches different from ours.

\section{Magnetized black holes}

\subsection{Kerr-Newman black holes in an external field}

The spacetime geometry describing a black hole immersed in an external backreacting magnetic field $B$ is given by the Ernst-Wild solution to the Einstein-Maxwell equations \cite{Ernst, ErnstWild, Wild}, which might be compared with the solutions in the linear approximation \cite{Wald, King}, cf. \cite{Bicak, Bicak2}. The full backreacting solution is characterized by three parameters, $m$, $a$, and $q$, which are related in an intricate way with the mass, the angular momentum, and the electric charge. While the solution can be thought of as a Kerr-Newman black hole embedded in a magnetic Melvin universe \cite{Melvin}, so that when $B=0$ the parameters $m$, $a$ and $q$ do agree with the mass, the angular momentum per unit of mass and the electric charge, respectively, when $B\neq 0 $, because of the backreaction, the relation between the three parameters and the {\it physical} conserved charges is more involved -- non-linear --. The precise relation has been debated in the literature, specially in connection to the mass and the first law of the black hole thermodynamics, cf. \cite{Gibbons, Gibbons2, Booth, Astorino, AstorinoKerrCFT, Lucho}; here we will contribute to that discussion.

Let us first consider the case $q=0$. We do this for the following reasons: First, while the full solution with three non-vanishing parameters can be written down analytically, the expression is more cumbersome and makes it more difficult to visualize the geometry. Second, in our previous paper \cite{Lucho} we considered the case $a=0$ with $q\neq 0$ in Eddington-Finkelstein type coordinates similar to those we will consider here, so that the expressions for the near horizon geometry can easily be found there. Third, the full expression in Boyer-Lindquist type coordinates can also be found in the original papers \cite{Ernst, ErnstWild}. Fourth, we will introduce the parameter $q$ later, in Section 3, as it is crucial to investigate the Meissner effect.

The solution with $m \neq 0 \neq a$ and $q=0$ in Boyer-Lindquist coordinates takes the form
\begin{equation}
    ds^2 = \Lambda (r,\theta)\, R^2(r,\theta) \Big( -  \frac{f(r)}{\Sigma (r,\theta) }\, dt^2 + \frac{dr^2}{f(r)} + d\theta^2  \Big) + \frac{\Sigma (r,\theta) \sin^2\theta }{\Lambda(r,\theta) R^2(r,\theta)}  \, ( d\phi - \omega(r,\theta)\,  dt)^2\label{Uno}
\end{equation}
with the metric functions
\begin{equation}
    f(r) = r^2 + a^2 - 2mr \quad , \quad R^2(r,\theta ) = r^2 +a^2 \cos^2\theta \quad , \quad \Sigma (r,\theta ) = (r^2+a^2)^2 - a^2 f(r)\, \sin^2\theta \nonumber 
\end{equation}
along with 
\begin{align}
    &\Lambda (r,\theta ) = 1 + \frac{B^2 \sin^2\theta }{2R^2(r,\theta )} \Big((r^2+a^2)^2 - a^2 f(r)\, \sin^2\theta \Big) + \frac{B^4}{R^2(r,\theta )} \Big[ \frac{R^2(r,\theta ) \sin^4\theta }{16} (r^2+a^2)^2  \nonumber \\
    & \ \ \ \ \ \ \ \ \ \ +  \frac{m a^2 r}{4} (r^2+a^2) \sin^6\theta + \frac{m^2 a^2}{4} \Big( r^2 (\cos^2\theta - 3)^2 \cos^2\theta + a^2(1+\cos^2\theta )^2 \Big) \Big]\, , \nonumber \\
    &\omega (r,\theta )= \frac{2mra}{\Sigma (r,\theta )} + \frac{B^4}{\Sigma (r,\theta )} \Big[\frac{a^3}{2} m^3 r (3+\cos^4\theta ) + \frac{am^2}{4} \Big( r^4 (3-6\cos^2\theta + \cos^4\theta ) 
      \\
    & \ \ \ \ \ \ \ \ \ \ + 2a^2 r^2 (3 \sin^2\theta - 2\cos^4\theta )- a^4(1+\cos^4\theta ) \Big) +\frac{amr}{8} (r^2+a^2)  \nonumber \\
    & \ \ \ \ \ \ \ \ \ \ \ 
    \times \, \Big( r^2 (3+6\cos^2\theta - \cos^4\theta ) - a^2 (1-6\cos^2\theta - 3\cos^4\theta )\Big) \Big]\, ;\nonumber
\end{align}
here, $t\in \mathbb{R}, \, r\in \mathbb{R}_{\neq 0 } $, and $\phi , \, \theta $ are two angular variables that chart the constant-$t$ surfaces of the event horizon. $B$, $m$ and $a$ are integration constants; the fourth integration constant, $q$, will be introduced latter. We will denote $r_0$ the radial location of the black hole event horizon, which exists provided $m^2\geq a^2$ (the condition for the existence of the horizon in the case the parameter $q$ is included reads $m^2\geq a^2+q^2$; see section 3). This solution is usually referred to as the Kerr-Newman-Melvin black hole, or the Kerr-Newman black hole in a Melvin universe. It is worth pointing out that, due to the $B$-dependent non-linear relation between the physical charges of the black hole and the parameters appearing in the metric, it turns out that even for $q=0$ the solution above describes an electrically charged rotating black hole. In fact, a specific relation between $m$, $a$, $B$ and $q$ is necessary for the solution to describe a spinning neutral black hole (see (\ref{LaQarga}) below). Melvin universe \cite{Melvin} corresponds to $m=a=q=0$ -- with $B$ being the external field that fills all space--, while the Kerr-Newman solution is obtained when $B=0$. We use units $G=c=1$.

The solution of the electromagnetic field reads
\begin{equation}
    A = [\Phi_0(r,\theta)\, -\, \omega (r,\theta)\, \Phi_3(r,\theta)]\, dt + \Phi_3(r,\theta)\, d\phi
\end{equation}
with
\begin{equation}
    \begin{split}
       \Phi_0 =& -\frac{a B^3}{8\Sigma (r,\theta)} \Big[4 a^4 m^2 + 2 a^4 m r - 24 a^2 m^2 r (m + r) - 6 m r^5 - 6r f(r)\, 2m (r^2+a^2) \cos^2\theta \\
       &- 4 a^2 m r^3 - 12m^2 r^4 + f(r)\, (2mr^3 + 4a^2 m^2 - 6a^2 mr)\cos^4\theta \Big]\, , \\ 
       \Phi_3 = & \frac{B}{R^2(r,\theta) \Lambda (r,\theta)} \Big[ \frac{\Sigma (r,\theta)}{2} \sin^2\theta + B^2 \Big( \frac{a^2}{2} m^2[ r^2 (3-\cos^2\theta )^2 \cos^2\theta + a^2(1+\cos^2 \theta )^2]  \\
       &+ \frac{a^2}{2} mr (r^2 + a^2) \sin^6\theta + \frac{R^2(r,\theta) }{8} (r^2 + a^2)^2 \sin^4\theta \Big) \Big]
    \end{split}\label{Omega}
\end{equation}
We emphasize that the solution is fully backreacting, so that it corresponds to an exact electrovacuum solution to the Einstein-Maxwell equations. The geometry of it has been extensively analyzed in the literature as well as its thermodynamics properties. For the latter, we refer to the relatively recent paper \cite{Astorino}. The geometry exhibits special features, such as non-compact ergoregions \cite{Gibbons} and horizons with sections of non-constant curvature, among others. It is related to other well-known solutions to Einstein-Maxwell theory, and it can also be generalized, for example, by including dyonic charges \cite{Gibbons} and acceleration \cite{Anabalon}. It is also related to interesting solutions in higher-dimensions \cite{Emparan}. 

\subsection{Near horizon limit}

Now, let us study the near horizon limit of the solution (\ref{Uno})-(\ref{Omega}). Our goal is to express the spacetime geometry and the gauge field configuration in a system of coordinates as the one introduced in \cite{Laura, Extended}. This would enable us to show that, near their vicinity, spinning magnetized black holes exhibit infinite-dimensional symmetries. More precisely, if we could prove that the Ernst-Wild solution can be written in the Eddington-Finkelstein type coordinates introduced in \cite{Laura, Extended} to perform the near horizon expansion, then we would {\it ipso facto} prove that there exist an infinite-dimensional isometry group that preserves the near horizon boundary conditions for these black holes.

Since $\omega(r,\theta)$ does not vanish at $r=r_0$, the first step to achieve our goal is to consider a boost $d\phi \rightarrow d\phi + c\, dt$ to produce a shift $\omega(r,\theta) \to \tilde{\omega }(r,\theta)=\omega(r,\theta) - \omega(r_0)$, with
\begin{equation}
    \omega(r_0) \equiv \omega_0 = \frac{a}{2 m r_0} + \frac{B^4 a}{8 r_0} \Big(3 r_0^3 + 3 a^2 r_0 + 2 a^2 m\Big)
\end{equation}
being a constant. This suffices to reach a comoving frame and make the angular velocity to be zero at the horizon.

Next, we perform the change of coordinates
\begin{align}
  v \, =\,  & t + \int  \frac{dr'}{f(r')} \, {\sqrt{\Sigma(r',\theta)}} \label{v_MKN}\\ 
  \varphi \, =\,  & \phi - \omega_0\, v + \int_{r_0}^{r} \frac{dr'}{f(r')} {(\omega(r',\theta) - \omega_0) \sqrt{\Sigma(r', \theta)} } 
  \label{phi_MKN}
\end{align}
which yields
\begin{align}
    &dv\, =\, dt + \frac{\sqrt{\Sigma (r,\theta )}}{f(r)} dr + \gamma(r,\theta ) d\theta\\
    &d\varphi \, =\, d\phi - \omega_0\, dv + (\omega(r,\theta )\, - \omega_0) \frac{\sqrt{\Sigma (r,\theta )}}{f(r)} dr + h(r,\theta)\, d\theta 
\end{align}
with
\begin{equation}
\gamma(r,\theta) = \int_{r_0}^r \frac{dr'}{f(r')}\, {\partial_{\theta}\sqrt{\Sigma(r',\theta)}} , \quad h(r,\theta) =  \int_{r_0}^r \frac{dr'}{f(r')} \, {\partial_{\theta}\Big((\omega(r',\theta) - \omega_0) \sqrt{\Sigma (r',\theta)}\Big)}
\end{equation}
With this, the metric takes the form
\begin{equation}
\begin{split}
    &ds^2 = \Lambda (r,\theta) R^2(r,\theta) \Big[ - \frac{f(r)}{\Sigma (r,\theta)} \, dv^2 + \Big( 1 -  \frac{\gamma^2(r,\theta) f(r)}{\Sigma (r,\theta)} \Big) d\theta^2 + \frac{2}{\sqrt{\Sigma (r,\theta)}} dr dv  
    \\
    & \ \ \ 
   + 2 \frac{\gamma (r,\theta) f(r)}{\Sigma (r,\theta)} dv d\theta  + 2 \frac{\gamma (r,\theta)}{\sqrt{\Sigma (r,\theta)}} dr d\theta \Big] + \frac{\Sigma (r,\theta) \sin^2\theta }{\Lambda (r,\theta) R^2(r,\theta)} \Big[ d\varphi - (\omega (r,\theta) - \omega_0) dv \\
    & \ \ \ + ((\omega (r,\theta) - \omega_0) \gamma (r,\theta) - h (r,\theta)) d\theta \Big]^2
\end{split}
\end{equation}
%\begin{equation}
%\begin{split}
%    &dS^2 = \Lambda \frac{f(r) R^2}{\Sigma} dv^2 + 2\Big[ \Lambda R^2\frac{\sqrt{\Sigma_0}}{\Sigma(r,\theta)} - \frac{\sqrt{\Sigma_0}}{f(r)} \frac{\Sigma}{R^2\Lambda} \sin^2(\theta) (\omega - \omega_0)^2 \Big] dv dr +\Lambda(r,\theta) R^2d\theta^2\\
%    &+ \Big[\frac{- a^2 \sin^2(\theta)}{\Sigma} \Lambda R^2 + \frac{\Sigma_0}{f(r)^2} \frac{\Sigma}{R^2\Lambda} \sin^2(\theta) (\omega - \omega_0)^2 \Big] dr^2 +\frac{\Sigma}{R^2\Lambda} \sin^2(\theta) \big{(}d\varphi - (\omega - \omega_0) dv  \big{)}^2\\
%    &+ 2 \frac{\sqrt{\Sigma_0}}{f(r)} \frac{\Sigma}{R^2\Lambda} \sin^2(\theta) (\omega - \omega_0) dr d\varphi,
%\end{split}
%\end{equation}
This coordinate system is regular at the horizon, on which it defines constant-$v$ slices with restricted metric
\begin{equation}
    ds^2 _{{|}_{H}} = \Lambda_0(\theta) R_0(\theta )^2 d\theta^2 + \frac{\Sigma_0 \sin^2\theta }{R_0^2(\theta )\Lambda_0(\theta)}  d\varphi^2\, ,
\end{equation}
where we denoted $R_0(\theta)=R(r_0,\theta)$, $\Lambda_0(\theta)=\Lambda(r_0,\theta)$ and $\Sigma_0=\Sigma (r_0,\theta)$. 
On $H$, we find the null vector $\ell = \partial_{v}$ and we can thus look for a vector $n$, also null, normalized such as $n_{\mu } \ell^{\mu } = 1$; namely,
%\begin{equation}\label{vectores}
%    n^{\mu} = \Big(0, -\frac{\sqrt{\Sigma_0}}{\Lambda_0 R_0^2}, 0, 0 \Big)^{\mu}
%\end{equation}
This is
\begin{equation}\label{vectores}
    n = \frac{\sqrt{\Sigma_0}}{\Lambda_0 (\theta ) R_0^2 (\theta ) } \partial_r
\end{equation}

Following the construction in \cite{Moncrief, Booth2}, we can consider a family of geodesics that cross $H$ with $n^{\mu}$ being the tangent vector, and consider that the geodesics are parameterized with an affine parameter $\rho$ such that $\rho_{|H} = 0$. Up to order $\mathcal{O}(\rho^2)$, this congruence of curves defines the vector field
\begin{equation}
    \Xi^{\mu }(v,\theta,\varphi, \rho) = \{v, r_0,\theta,\varphi \}^{\mu } + \rho \, n^{\mu } + \frac{\rho^2}{2} {\partial_{\rho }^2 \Xi^{\mu }}_{{|}_{\rho = 0}} + \mathcal{O}(\rho^3)
\end{equation}
with the second derivative being defined by the geodesic equation $
 {\partial_{\rho }^2 \Xi^{\alpha}}_{{|}_{\rho = 0}} = - \Gamma_{\mu \nu}^{\alpha} n^{\mu} n^{\nu}$. Given the way in which $n$ has been defined, we have the following gauge conditions for the radial components of the spacetime metric $g_{\rho v}= n _{\mu} \ell^{\mu}= 1, \,  g_{\rho \rho}= n _{\mu} n^{\mu} =0,  g_{\rho A}=n_{\mu} e_{A}^{\mu} =0$, with $A=1,2$ referring to the coordinates on the constant-$v$ slices of the horizon; we will often use the notation $z^A=\{ \varphi \, , \, \theta \}$ to refer to the angular coordinates.

To obtain the other metric components up to order $\mathcal{O}(\rho )$, it is sufficient to consider the variation, up to that order, in the direction generated by the affine parameter $n=\partial_{\rho} $; namely
\begin{equation}
    g_{\mu \nu} = g_{\mu \nu}^{(0)} +    g_{\mu \nu}^{(1)}\, \rho + \mathcal{O}(\rho^2) 
\end{equation}
with $g_{\mu \nu}^{(0)} = g_{\mu \nu\, |\rho = 0}$ and $ g_{\mu \nu}^{(1)} = ( \mathcal{L}_{n}  g )_{\mu \nu\, | \rho = 0}$. Let us be reminded that the horizon, $H$, which is located at $r=r_0$, in these new coordinates would be $\rho =0$. At order $\mathcal{O}(\rho )$, the non-vanishing components of the metric are
\begin{equation}
\begin{split}
    &g_{v v}^{(1)} = -\frac{f'(r_0)}{\Sigma_0^{1/2}}\\
    &g_{v \theta}^{(1)} =- \frac{1}{\Lambda_0 (\theta ) R_0^2(\theta )} \partial_{\theta} \big( \Lambda R^2 \big) _{|r=r_0}\\
    &g_{\theta \theta}^{(1)} =  \frac{\Sigma_0^{1/2}}{\Lambda_0(\theta ) R_0^2(\theta )}\, \partial_{r} \big( \Lambda R^2 \big)_{|r=r_0}\\
     &g_{v \varphi}^{(1)} = -\frac{\Sigma_0^{3/2} \sin^2\theta }{\Lambda_0^2(\theta ) R_0^4(\theta )} \, (\partial_{r}  \omega )_{|r=r_0}\\
     &g_{\varphi \varphi}^{(1)} =  \frac{\Sigma_0^{1/2} \sin^2\theta}{\Lambda_0(\theta ) R_0^2(\theta )}  \, \partial_{r} \Big( {\Sigma}{\Lambda^{-1} R^{-2}} \Big) _{|r=r_0}\\
     &g_{\theta \varphi}^{(1)} = -\frac{\Sigma_0^{2} \sin^2\theta }{\Lambda_0^2(\theta ) R_0^4(\theta )}\, \partial_{\theta}  \Big({\omega}{f}^{-1} \Big) _{|r=r_0}\, ;
\end{split}
\end{equation}
and then we have $g_{vv}^{(0)}= g_{v\varphi }^{(0)} = g_{v\theta }^{(0)}=0$, $g_{\rho v }^{(0)}=1$ , as well as $g_{\rho \alpha }^{(0)} = 0$ with $\alpha =1,2,3$. $\kappa =-\frac 12 g^{(1)}_{v v }$ is the surface gravity at the horizon, ultimately associated to the Hawking temperature $T= \kappa/(2\pi )$ ($k_B=\hbar =1$). -- For the extremal configuration the surface gravity vanishes ($\kappa = 0$), and in that case the analysis of the charges has to be done separately.-- In the new coordinate system, the metric takes a form that satisfies the asymptotic boundary conditions at the horizon considered in \cite{Laura, Extended}. This means that the magnetized black hole geometry admits infinite asymptotic Killing vectors preserving the near horizon form. The next step would be to verify whether the gauge field also admits the correct asymptotic conditions, cf. \cite{Pu, Puhm}. In order to check that, let us express the electromagnetic potential $A = A_{t} \, dt + A_{\phi} \, d\phi$ in the coordinates introduced in (\ref{v_MKN})-(\ref{phi_MKN}). This yields
\begin{align}
    \nonumber
    A = A_{t}  \hspace{0.2em} dv - \frac{\sqrt{\Sigma (r, \theta )}}{f(r)} [A_{t} + (\omega (r, \theta ) - \omega_0) A_{\varphi})]\hspace{0.2em} dr + A_{\varphi} \hspace{0.2em} d\varphi - A_{\varphi} h (r, \theta ) \hspace{0.2em} d\theta
\end{align}
Next, we can use a residual gauge freedom to make $A_{\rho} = 0$ at $H$. That is, we perform the gauge transformation $A \rightarrow A + d\zeta$ with
\begin{equation}
    \zeta = \int_{r_0}^{r} {dr'} \frac{\sqrt{\Sigma(r',\theta)}}{f(r')} \Big[ A_{t}(r',\theta) + (\omega (r',\theta)- \omega_0) A_{\varphi}(r',\theta)\Big]
\end{equation}
which allows to write
\begin{equation}
    A = [\Phi_0 (r,\theta) - (\omega (r,\theta)-\omega_0) \Phi_3 (r,\theta) ]  \hspace{0.2em} dv + \Phi_3 (r,\theta) \hspace{0.2em} d\varphi + [\Phi_3 (r,\theta) h(r,\theta) - \partial_{\theta}\zeta] \hspace{0.2em} d\theta\label{La20}   
\end{equation}
This expression does satisfy the right asymptotic conditions for the electromagnetic field at the horizon; namely,
\begin{align}\label{potencialA}
    &A_{v} = A_{v}^{(0)} + \rho \hspace{0.2em} A_{v}^{(1)}(v,z^{A}) + \mathcal{O}(\rho^2)\\
    \nonumber
    &A_{B} = A_{B}^{(0)}(z^{A}) + \rho \hspace{0.2em} A_{B}^{(1)}(v,z^{A}) + \mathcal{O}(\rho^2)
\end{align}
where $A_{v}^{(0)}$ is a fixed constant, $A_{B}^{(0)}$ with $B=1,2$ only depend on the angular variables $z^B=\{\varphi , \theta \}$, and $A_{\rho} = 0$. To obtain the potential at order $\mathcal{O}(\rho )$, we follow a similar procedure as before: we expand the electromagnetic potential around $r\simeq r_0$ as follows $A_{\mu} = A_{\mu}^{(0)} + \rho  A_{\mu}^{(1)} + \mathcal{O}(\rho^2)$, where $A_{\mu}^{(0)} = A_{\mu\, |H}$ and $ A_{\mu}^{(1)} = (\mathcal{L}_{n} \, A )_{\mu\, |H}$; see (\ref{LaBelow}) below. In this way, we obtain
\begin{equation}
    \begin{split}
        &A_{v}^{(0)} = \Phi_0(r_0, \theta )\\ 
        &A_{\varphi}^{(0)} = \Phi_3(r_0, \theta ) - \Phi_3(r_0, 0 )\\
        &A_{\theta}^{(0)} = 0
    \end{split}
\end{equation}
along with
\begin{equation}
    \begin{split}
        &A_{v}^{(1)} =- \frac{{\Sigma_0^{1/2}}}{\Lambda_0(\theta )  R_0^2(\theta ) } \Big( \Phi_3 \, \partial_r \omega - \partial_r \Phi_0 \Big) _{|r=r_0}\\
        & A_{\varphi}^{(1)} =  \, \, \frac{{\Sigma_0^{1/2}}}{\Lambda_0(\theta )  R_0^2(\theta ) } (\partial_r \Phi_3) _{|r=r_0} \\
        & A_{\theta}^{(1)} = -\frac{{\Sigma_0^{1/2}}}{\Lambda_0(\theta )  R_0^2(\theta ) } \Big(  \partial_r \partial_{\theta} \zeta  - \Phi_3 \, \partial_r h \Big) _{|r=r_0}
    \end{split}
\end{equation}
where we have added a constant to $A_{\varphi}$ using the remnant gauge freedom; with this, the potential vaishes both at the north and the south pole. In this way, we have explicitly shown that the magnetized black hole geometry obeys the horizon boundary conditions discussed in \cite{Laura, Extended, Puhm}, and therefore the magnetized horizon of (\ref{Uno}) enjoys asymptotic infinite-dimensional supertranslation and superrotation symmetries.

\subsection{Near horizon symmetries}

In the next section we will consider the Noether charges associated to the infinite-dimensional symmetries we just discussed. In preparation to do so, let us review the form of the asymptotic Killing vectors and gauge transformations that generate such symmetries: The near horizon expansion discussed above corresponds to the expansion around $r\simeq r_0$,  i.e. $\rho \simeq 0$, in powers of $\rho $, and the diffeomorphisms and gauge transformations that preserve such asymptotia are known to be of the form
\begin{equation}
\delta g_{\mu \nu} = \mathcal{L}_{\chi }g_{\mu \nu } \, , \ \ \ \ \delta A_{\mu} = \mathcal{L}_{\chi } +\partial_{\mu} \epsilon\label{LaBelow}
\end{equation}
with 
\begin{equation}
\chi ^v = T(z^A) +\mathcal{O}(\rho )\, , \ \ \
\chi ^{\rho } = \mathcal{O}(\rho )\, , \ \ \
\chi^A = Y^A (z^B) +\mathcal{O}(\rho )\, ,
\end{equation}
and
\begin{equation}
\epsilon = U(z^A) -T(z^A)A_v^{(0)} +\mathcal{O}(\rho ) \, ,\label{NoB}
\end{equation}
where $T(\varphi , \theta ), \, Y^{\varphi }(\varphi , \theta ), \, Y^{\theta }(\varphi , \theta ),$ and $ U(\varphi , \theta )$ are arbitrary constants of the angular coordinates. The Fourier modes of the expansion of these functions in the angular variables $z^A$ on $H$ generate an infinite-dimensional current algebra in semidirect sum with another set of supertranslations and two copies of Witt algebra; see \cite{Extended} for details. In the next section, we construct the charges associated to the symmetries generated by (\ref{LaBelow})-(\ref{NoB}).

\section{Conserved charges}
\subsection{Noether charges on the horizon}

The symmetries discussed above have associated the following Noether charges \cite{Glenn, Glenn2}
\begin{equation}
Q[T,Y^A, U] = -\frac{1}{16\pi }\int_{H} dS \Big( T\, g_{vv}^{(1)}+Y^A (g_{vA}^{(1)} + 4A_{A}^{(0)} A_v^{(1)} )+4U\, A_v^{(1)} \Big)\label{LACARGASSS}
\end{equation}
where 
\begin{equation}
dS= \sqrt{\det g_{AB}^{(0)}}\, d\varphi d\theta \, 
\end{equation}
is the measure of the constant-$v$ slices on $H$; we will occasionally write $d\varphi d\theta=d^2z$. The subindex $H$ in the integral in (\ref{LACARGASSS}) refer to constant-$v$ slices on $H$. These are charges computed at the horizon and are defined by integrating on the constant-$v$ slices. The values associated to the zero modes are
\begin{equation}
S=\frac{2\pi }{ \kappa}Q[1,0,0]  \, , \ \ \ \ j=Q[0,\delta^{A}_{\varphi }, 0]  \, , \ \ \ \  e=Q[0,0,1] \, ;
\end{equation}
they are the entropy, the angular momentum and the electric charge, respectively. While entropy (in the non-extremal case) is associated to rigid translations $\partial_v$ on $H$, the angular momentum is the charge associated to $\partial_{\varphi }$ defined also on $H$.

\subsection{Angular momentum and Komar integrals}

While the analysis performed above is valid in general, the explicit expressions we wrote in Section 2 correspond to the particular case $q=0$. Now, let us consider the most general expressions. The explicit solution with arbitrary parameters $m$, $a$, $q$ and $B$ can be found in \cite{Ernst, ErnstWild, Gibbons, Gibbons2, Booth, Astorino}, and the near horizon analysis with $q\neq 0$ was done in \cite{Lucho} for the case $a=0$. As we will see, the total angular momentum explicitly depends on $q$ and $B$, and not only on $a$. When $a\neq 0 \neq q$ the metric functions depend on both parameters and, as in the Kerr-Newman solution, the horizon location $r_0$ depends, not only on the mass, but also on both the angular momentum and the electric charge: $f(r)=r^2+a^2-2mr+q^2$, so that $r_0=m^2+ \sqrt{m^2-a^2-q^2}$.

Computing the charge $Q[0,\delta^{A}_{\varphi }, 0]$, which corresponds to the angular momentum, amounts to calculate the integral
\begin{equation}\label{J}
    j = -\frac{1}{16 \pi}\int_{H} dS\, \big(g_{v \varphi}^{(1)} + 4 A_{\varphi}^{(0)} A_{v}^{(1)} \big) 
\end{equation}
on the horizon. Despite being a concrete analytic expression, in the case of the magnetized horizon the evaluation of (\ref{J}) is quite cumbersome; see \cite{Lucho} for the explicit computation in the case $q\neq0=a$. What we will rather do here is to prove that (\ref{J}) admits to be expressed as a Komar integral on $H$; this will allow us to compare with the results in the literature. In order to do so, let us separate (\ref{J}) in two contributions: a first contribution coming from the spacetime geometry, which corresponds to the first term in the integrand, and a second contribution coming from the electromagnetic field, which corresponds to the second term in the integrand. Each of these contributions will be shown to match the corresponding Komar expression. As for the first one, it is possible to show that it matches the integral formula
\begin{equation}
    J_{K} = \frac{1}{16 \pi} \int_{{}} *\, dK
\end{equation}
with $K^{\mu}$ being the rotational Killing vector $\partial_{\varphi}$ and $dK $ stands for the exterior derivative of $K_{\mu}= g_{\mu \varphi}$, namely
\begin{equation}
    dK = \partial_{\rho}g_{v \varphi}\, d\rho \wedge dv + \partial_{\theta}g_{v \varphi} \,  d\theta \wedge dv + \partial_{\theta}g_{\varphi \varphi}\,  d\theta \wedge d\varphi +
    \partial_{\rho}g_{\theta \varphi}\, d\rho \wedge d\theta + \partial_{\rho}g_{\varphi \varphi}\, d\rho \wedge d\varphi
\end{equation}
with Hodge dual $*dK_{\mu \nu  } = \frac{1}{2} \sqrt{- g}\, dK^{\alpha \beta} \epsilon_{\alpha \beta \nu \mu  }$, whose explicit expressions can be found, for example, in \cite{Booth}. To compare with our expression (\ref{J}), we evaluate the component $*dK_{\theta \varphi }$ on the horizon, namely
\begin{equation}
    *dK_{\theta \varphi \, |H} = \sqrt{\text{det} g^{(0)}_{AB}}\,  \epsilon_{v \rho \theta \varphi}\, dK^{v \rho}_{|H} = - \sqrt{\text{det} g^{(0)}_{AB}}\, \partial_{\rho}(g_{v \varphi})_{|H}\, ,
\end{equation}
and, by expanding in $\rho $, we get
\begin{equation}
    J_{K} = -\frac{1}{16 \pi} \int_{{H}} dS \, g_{v \varphi}^{(1)} \, .
\end{equation}
That is to say, the first contribution to the horizon charge (\ref{J}) is found to agree with a Komar integral. Now, let us consider the second contribution. For asymptotically flat spacetimes the entire contribution to the angular momentum would be $j=J_K$; however, in the Melvin universe the gauge field configuration does not vanish at infinity and the second term in (\ref{J}) does contribute to the Komar integral; it does with a term \cite{Booth}
\begin{equation}
    J_{EM} = \int_{{H}} dS\, K^{\alpha} \mathcal{J}_{\alpha} \qquad \text{with} \qquad \mathcal{J}_{\alpha} = \frac{1}{4\pi} \ell^{\mu} n^{\nu} F_{\mu \nu} (g^{\beta}_{\alpha} + \ell^{\beta} n_{\alpha} + \ell_{\alpha} n^{\beta}) A_{\beta}
\end{equation}
where now $K^{\alpha }$ is a rotational Killing vector, $\ell$ and $n$ are the two transversal null vectors on $H$, and the integrand $E_{\perp} \equiv \ell^{\mu} n^{\nu} F_{\mu \nu}$ is the transversal component of the electric field. Evaluating this expression on $H$, where we can use $\ell = \partial_{v}$ and $n = \partial_{\rho}$, we obtain
\begin{equation}
    J_{EM} = \frac{1}{4\pi}\int_{{H}}  dS\, (F_{v \rho} A_{\varphi}) _{|H} = - \frac{1}{4\pi}\int_{{H}} dS\, A^{(1)}_{v} A^{(0)}_{\varphi} \, ,
\end{equation}
which exactly reproduces the second term in (\ref{J}). In other words, we have shown that the contribution to the horizon charge (\ref{LACARGASSS}) that corresponds to the angular momentum admits to be written as Komar integrals on the horizon; namely
\begin{equation}
Q[0,\delta^{A}_{\varphi }, 0]= J_{K} + J_{EM} \, .
\end{equation}
%Therefore, we have shown that the horizon charge $Q[0,\delta^{A}_{\varphi }, 0]$ agree with the full Komar integral on the horizon.

In \cite{Lucho}, the angular momentum of the black hole immersed in a magnetic Melvin universe was computed from the near horizon perspective for the case $q\neq a=0$, resulting in
\begin{equation}
j_{|a=0}=-q^3B(1+\frac 14 q^2B^2)\, .
\end{equation}
This result was found to be consistent with the angular momentum computed by other methods in \cite{Gibbons, Gibbons2, Astorino, Booth, AstorinoKerrCFT}, which in the general case reads 
\begin{equation}
j= am-q^3B-\frac 32 amq^2B^2 - (2qa^2m^2+\frac 14 q^5)B^3-(a^3m^3+\frac{3}{16}q^4am)B^4\, 
 .\label{LaJota}
\end{equation}
The angular momentum results to be a finite expansion in powers of $B$, which sometimes it is convenient to write as a polynomial in $am$ or as a polynomial in $q$.
%Namely, 
%\begin{equation}
%j= -\Big(q^3B+\frac 14 q^5B^3 \Big) + \Big(1-\frac 32 q^2B^2 -\frac{3}{16} q^4B^4 \Big)\, a m  - 2qB^3\,  a^2 m^2  - B^4\,  a^3 m^3  \label{LaJota2}
%\end{equation}
We notice from (\ref{LaJota}) that when $B=0$ the angular momentum reduces to the standard result $j=ma$ of Kerr-Newman black holes. Also, we notice that when $q=0$ the angular momentum receives a contribution from the external magnetic field, yielding $j=am(1-a^2m^2B^4)$. When $a=0$ the angular momentum is $j=-q^3B(1+\frac 14 q^2B^2)$. It is also worth noticing that the parity and charge conjugation symmetry express themselves in the fact that expression (\ref{LaJota}) is invariant under the transformations $\{ a\to -a,\, j\to -j,\, q\to \mp q ,\, B\to \pm B\}$ and under the transformation $\{q\to - q ,\, B\to - B \}$.

\subsection{Wald entropy}

Now, we can analyze the other conserved charges, one of them corresponding to the black hole entropy. As shown by Wald \cite{WaldEntropy}, the black hole entropy can be expressed as a Noether charge computed at the horizon. It was observed in \cite{Laura} that one of the charges (\ref{LACARGASSS}), the one corresponding to the zero mode $T=1$, which realizes rigid translations in the coordinate $v$, actually reproduces the Wald entropy charge. In other words, the charge associated to the Killing vector $\chi = \partial_v$ gives the Bekenstein-Hawking entropy formula multiplied by the Hawking temperature, namely
\begin{equation}
Q[1,0,0]=-\frac{1 }{16\pi }\int_H dS\, g_{vv}^{(1)}= \frac{ \kappa }{2\pi }\, \frac{A}{4 } = TS\, ,\label{entro}
\end{equation}
($G=c= k_B =1$), with $A$ being the area of the horizon. The second equality in (\ref{entro}) simply follows from $g_{vv}^{(1)}$ being constant. When evaluating the component $g_{vv}^{(1)}$ above, it is worth noticing that, when $q\neq 0$, $f'(r_0)=2\sqrt{m^2-a^2-q^2}$.

\subsection{Electric charge and Gauss law}

Now, let us move to the electric charge. The claim is that it corresponds to 
\begin{equation}\label{sec3eq0}
    Q[0,0,1] = - \frac{1}{4 \pi}\int_H dS\, A_{v}^{(1)}\, ,
\end{equation}
evaluated on $H$. In order to prove that this gives the correct result, we can compare with the canonical form, namely with the computation of the electric charge as the integral of the dual 2-form $*F$ over a 2-dimensional surface that encloses the black hole at a distance. In simple words, {the Gauss law in the black hole background} should give us the total electric charge of the system
\begin{equation}\label{sec3eq1}
    e = \frac{1}{4 \pi}\int *F \, .
\end{equation}
What we will show here is that the integral representations (\ref{sec3eq0}) and (\ref{sec3eq1}) actually coincide. The strategy is simple: since the flux (\ref{sec3eq1}) can be taken over any 2-dimensional constant-$r$ and constant-$t$ surface, while in contrast (\ref{sec3eq0}) is defined as an integral on the constant-$v$ sections of $H$, we will first compute the charge (\ref{sec3eq1}) at fixed $r = r_{0}$. Moreover, we will work in the gauge where $\omega(r_{0},\theta) = 0$ to impose the required near horizon boundary conditions. In advanced coordinates and in the original gauge, the electromagnetic potential (\ref{La20}) has the form
\begin{equation}
    A = [\Phi_0 (r,\theta) - (\omega (r,\theta)-\omega_0) \Phi_3 (r,\theta) ]  \hspace{0.2em} dv + \Phi_3 (r,\theta) \hspace{0.2em} d\varphi + \Phi_3 (r,\theta) h(r,\theta)  \hspace{0.2em} d\theta \, ,   
\end{equation}
which in terms of the original Boyer-Lindquist type coordinates reads
\begin{equation}\label{sec3eq2}
    A = \Big[\Phi_0 (r,\theta ) - \omega (r,\theta ) \Phi_3 (r,\theta )\Big] dt + \Phi_3 (r,\theta )d\phi \equiv A_{t} \, dt + A_{\phi}\, d\phi \, ;
\end{equation}
this expression can be found in appendix B of \cite{Gibbons}, the angular frequency $\omega $ is shifted with respect to the function defined in Eq. (B.8) therein in order to fix the right boundary conditions. With this, we compute the field strength
\begin{align}\label{sec3eq3}
F_{\mu \nu } dx^{\mu } \wedge dx^{\nu } = \partial_{r}A_{t} \,  dr\wedge dt + \partial_{\theta}A_{t}\,  d\theta \wedge dt +  \partial_{r}A_{\varphi}\,  dr\wedge d\varphi +\partial_{\theta}A_{\varphi}\,  d\theta \wedge d\varphi
\end{align}
from which we get the components of its dual $*F_{\mu \nu } = \frac{1}{2} \sqrt{- g}F^{\alpha \beta} \epsilon_{\alpha \beta \mu \nu }$. Since we are interested in writing the integral (\ref{sec3eq1}) as the flux through a surface defined at constant $r$ and constant $t$, the only component we need to look at is $*F_{\theta \varphi} $. For the spacetime metric given by (\ref{Uno}) -- supplemented with the dependence on $q$--, after some algebra one finds that $*F_{\theta \varphi} $ evaluated at the horizon takes the form
\begin{equation}\label{sec3eq71}
    *F_{\theta \varphi \,|H } = -\epsilon_{t r \theta \varphi}  \,  \frac{\Sigma_0 \sin\theta}{\Lambda _0(\theta ) R_0^2(\theta )}  \Big( \partial_{r}A_{t}\Big) _{|r=r_0}
\end{equation}
which can be written as
\begin{equation}\label{sec3eq7}
    *F_{\theta \varphi \, |H} = \sqrt{\Sigma_0} \sin\theta \, \times \, \frac{\sqrt{\Sigma_0}}{\Lambda_0(\theta ) R_0^2(\theta )} \Big( \Phi_3 \partial_{r}\omega - \partial_{r}\Phi_0 \Big) _{|r=r_0}\, .
\end{equation}
It turns out that this Equation is both simple and remarkable: while the first factor in (\ref{sec3eq7}) is the square root of the determinant of the induced metric on the 2-dimensional surface, the second factor gives the correct contribution of the electric potential; namely
\begin{equation}
\sqrt{\Sigma_0} \sin\theta=\sqrt{\det g^{(0)}_{AB}} \, , \ \ \text{and}\  \ \ \frac{\sqrt{\Sigma_0}}{\Lambda_0(\theta ) R_0^2(\theta )} (  \partial_{r}A_{t})_{|r=r_0}=A_{v\, }^{(1)}.
\end{equation}
Therefore, the equivalence of the two methods is completely proven; we find
\begin{align}\label{sec3eq8}
    e &= \frac{1}{4 \pi}\int *F = - \frac{1}{4 \pi}\int_H dS\, A_{v}^{(1)} = Q[0,0,1] \, .
\end{align}

In terms of the black hole parameters, the electric charge takes the following form
\begin{equation}\label{LaQarga}
    e =  q (1-\frac 14 q^2B^2) +2amB
    %{-} [q(1 - \frac{1}{4} q^2 B^2 ) + 2amB]
\end{equation}
From this we observe that when $B=0$ the electric charge reduces to the standard result for the Kerr-Newman geometry, namely $e=q$. We also notice that when $B\neq 0$ the physical charge $e$ receives a contribution from both the spin and the external magnetic field; it exhibits a non-linear dependence with $q$ and it yields a finite value $e=2amB$ in the case $q=0$. Due to the non-linear term in $q$, when $a=0$ the electric charge can still be zero for non-vanishing values of $q$ provided the condition $q=\pm 2/B$ is satisfied. This yields a critical value $B_c$ for the field for which $e=0$ for $q\neq 0$, namely $|B_c|=2/|q|$; at this value $j$ does not necessarily vanish. It is also worth noticing that the symmetry under parity and charge conjugation express themselves in the fact that expression (\ref{LaQarga}) is invariant under the transformations $\{ q\to -q,\, e\to -e,\, a\to \mp a ,\, B\to \pm B\}$ and under the transformation $\{a\to - a ,\, B\to - B \}$.

In conclusion, we can say that from (\ref{LACARGASSS}) we obtained the right conserved charges of the charged spinning black hole immersed in a backreacting magnetic field. This analysis, however, was valid for the case of non-extremal black holes. As explained in \cite{Extended}, the first term in the expression (\ref{LACARGASSS}) for the charges gets modified in the extremal limit, which is the case we are mainly interested in -- as it is when the Meissner effect can occur--. The method to compute the horizon charges in the extremal case has been worked out in \cite{Puhm}, and in Section 4 we will discuss its application to analyze the black hole in the Meissner state.

\subsection{Thermodynamics of magnetized black holes}

Before moving to analyze the Meissner effect, let us review the thermodynamics in the case of magnetized black holes for generic values of $a$, $q$ and $m$. In the ensemble defined by keeping the external field fixed, the first law of black hole mechanics takes the form
\begin{equation}
dM=\frac{\kappa}{2\pi} dS +\Omega \,dj + \Phi\, de  \, ,\label{1spp}
\end{equation}
where the explicit expressions for the angular velocity $\Omega$ and the electric potential $\Phi$ at the horizon can be found in \cite{Astorino}. $M$ in (\ref{1spp}) is the Christodoulou-Ruffini mass, which takes a cumbersome expression in terms of the parameters $m,\, a,\, q$ and $B$. More precisely, $M^2$ is a polynomial of degree 4 in $B$ with coefficients that depend on $m,\, a$ and $q$; see Eq. (44) in \cite{Astorino}. In terms of this mass and the other physical charges, the constraint $m^2\geq a^2+q^2$, which is the condition for avoiding naked singularities, translates into
\begin{equation}
M^2\geq \frac 12 \Big( e^2+\sqrt{e^4+4j^2}\Big)\, ,
\end{equation}
which reduces to the standard inequalities $M\geq |e|$ and $M\geq |j|/M$ in the cases $j=0$ and $e=0$, respectively.

In addition to the first law (\ref{1spp}), the black hole quantities obey the Smarr type formula
\begin{equation}
M=\frac{\kappa}{\pi}S+2\Omega\, j + \Phi\, e\, ,
\end{equation}
which notably simplifies when the black hole is in the Meissner state $e=0$, $M^2=|j|$.

In \cite{Gibbons2}, the analysis of the thermodynamics of magnetized black hole was done by considering the variation of the external magnetic field $B$ in the ensemble. This leads to the definition of a gravitational energy $E$ that satisfies the following form for the first law of black hole mechanics
\begin{equation}
dE+\mu\, dB=\frac{\kappa}{2\pi} dS +\Omega \,dj + \Phi\, de  \, ,
\end{equation}
cf. (\ref{1spp}), with the magnetic momentum being
\begin{equation}
\mu =\frac{je}{2m}\, g \ \ \ \ \text{with} \ \ \ \ g-2\, =\, \mathcal{O}(j^2B^4)+\mathcal{O}(e).
\end{equation}
$g$ is gyromagnetic ratio, which is found to be $g\simeq 2$ up to general relativity corrections due to gravitational backreaction, in agreement with the original Carter's result \cite{Carter}.

\section{Meissner effect}

\subsection{The phenomenon}

Now, let us go back to the Meissner effect: By considering the solution of Maxwell equations describing a Kerr black hole in a background magnetic field, which was first studied by Wald in \cite{Wald} in the probe approximation, King et al. made in \cite{King} a remarkable observation: as it approaches extremality, spinning black holes expel the lines of magnetic field. More precisely, they found that the flux of magnetic field through the event horizon hemisphere decreases monotonically from $4\pi m^2B$ to zero as the angular momentum increases. In their own words, the lines of force of the magnetic field seem to experience a centrifugal repulsion as the hole spun up. This is the black hole Meissner effect, and it is 
%In the presence of electric charge, the phenomenon ceases, and this is related to the presence of currents in the spinning charged case; in contrast, the extremal uncharged black holes do exhibit this sort of superdiamagnetism as it was later 
confirmed by the analysis of the full backreacting solution \cite{backreaction, Voro, Budin}. In fact, the Meissner effect is a quite generic features of stationary axisymmetric solutions \cite{Bicak, Bicak2, Mac}, having been observed, for example, in charged black holes \cite{Mac, Bicak, Ruffini} and in extended solutions in higher-dimensions \cite{me11}. 

We will focus on uncharged black holes. The Meissner effect then takes place when the black hole is maximally rotating. According to our charge computation, the zero electric charge condition is $Q[0,0,1]=0$, which reads
\begin{equation}
a= \frac{1}{2m} \Big( \frac 14 q^3B -\frac qB \Big) \label{LaA}
\end{equation}
On the other hand, the extremality condition is
\begin{equation}
m=\sqrt{q^2+a^2}\, .\label{LaB}
\end{equation}
It is worth mentioning that the neutrality condition (\ref{LaA}) is valid for both the non-extremal and the extremal cases. While the charge computation performed above has been addressed for the non-extremal case and the near horizon charges in general receive a modification in the extremal case, the expression charge associated to the gauge field remains the same in both cases; see (\ref{Laquesigue}) below. The horizon charge that does get modified in the extremal case is the one associated to supertranslations in the $v$ direction, as we will discuss below.

One can explicitly verify that, provided both (\ref{LaA}) and (\ref{LaB}) are satisfied, the magnetic field vanishes at the horizon. To see this, we can take a look at the radial component magnetic field at the horizon, which is given by
\begin{equation}
B_{r|H}=\frac{1}{\sqrt{\Sigma_0 }\sin \theta }\, ({\partial }_{ \theta }\Phi_3 )_{|r=r_0}\, ,\label{Berre}
\end{equation}
and the azimuthal component of the magnetic field at the horizon, which is
\begin{equation}
B_{\theta |H}=\frac{\sqrt{f(r_0)}}{\sqrt{\Sigma_0 }\sin \theta }\, ({\partial }_{ r }\Phi_3 )_{|r=r_0}\, .\label{Btita}
\end{equation}
It turns out that, when both (\ref{LaA}) and (\ref{LaB}) hold, both (\ref{Berre}) and (\ref{Btita}) vanish; that is,
\begin{equation}
B_{r|H}=0\, , \ \ \ \ B_{\theta |H}=0\, .
\end{equation}
A particular case in which this happens is $m=\pm q$ with $q=\pm 2/B$, which yields $a=0$ and $|j|=4/|B|=2|q|$. However, this is a much more general phenomenon that occurs always that $e=0$ and $m^2=a^2+q^2$. We will focus on solutions that are continuously connected with the Kerr-Newman solution when $B\to 0$. For such solutions, the validity of the neutral condition (\ref{LaA}) and the extremality condition (\ref{LaB}) implies the following relation between the parameters $m$, $q$ and $B$
\begin{equation}
B_{\sigma }=\frac{4m\, \text{sign} (a)}{q^3} \Big( \sqrt{m^2-q^2}+\sigma  m-\sigma  \frac{q^2}{2m} \Big) \, ,
\end{equation}
with $\sigma =\pm 1$ indicating two different (Meissner) branches; cf. Eqs. (50)-(52) in \cite{Astorino}. Noticing that, for the extremal configuration, $\text{sign}(a)\sqrt{m^2-q^2}=a$, the value of the magnetic field in each branch can be written as
\begin{eqnarray}
B_{\sigma}  = {2\sigma \, \text{sign}(qa)  } \, \frac{(m+ \sigma |a|)^{ 1/2}} {(m- \sigma |a|)^{ 3/2}}\, . \label{Bsigma}
\end{eqnarray}
This expresses that, when $B=0$, the extremality condition for the neutral black hole reduces to $m=|a|$; and it also shows the existence of branches that are not continuously connected to the Kerr-Newman solution in the limit $B\to 0$. Of course, each branch ($\sigma =\pm 1$) of solutions of (\ref{Bsigma}) remains unchanged when changing $\{B_{\sigma },q,a\}\to \{-B_{\sigma },\pm q,\mp a\}$ or $\{B_{\sigma },q,a\}\to \{B_{\sigma },- q,- a\}$, as it follows from (\ref{LaA}).

\begin{figure}
\centering
\includegraphics[scale=0.50]{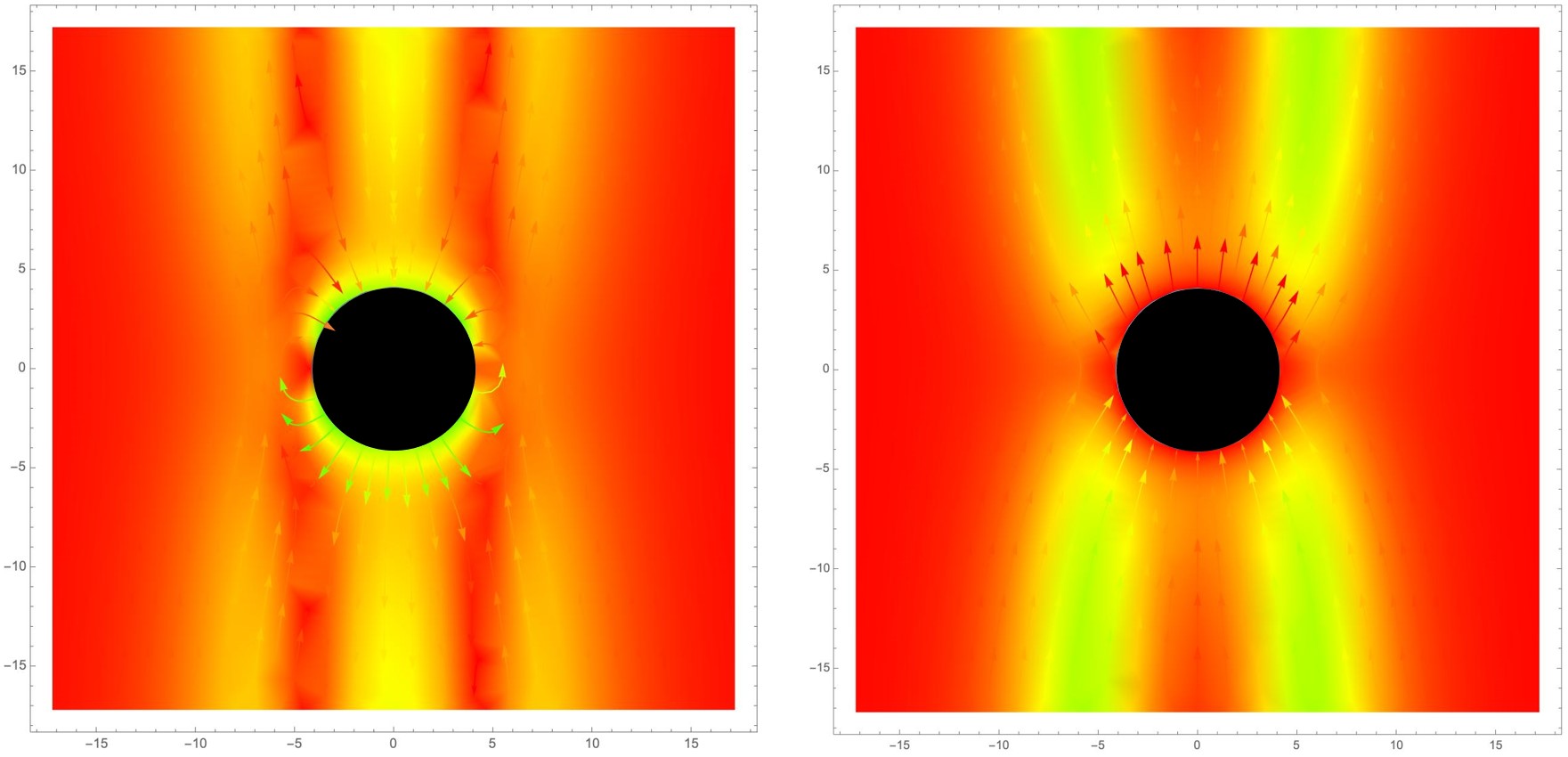}
\caption{Lines of magnetic field close to the event horizon for the case of a black hole exhibiting Meissner effect (right) and one that is not in its Meissner state (left). Red colour indicates the zones where the magnetic field is weaker, while yellow colour indicates where the magnetic field is stronger.  The parameters of the black hole in the Meissner state (right) are such that $|j|=M^2$ and $e=0$, while in the other configuration (left) $0<|j|<M^2$ and $|e|> 0$. The plots represent a transversal section of the solution, with the equatorial plane being perpendicular to the vertical axis. The plot is in Cartesian coordinates, having adimensionalized using $a=m$ with $B$ being a free parameter to control the configuration.} \label{fig:1}
\end{figure}

It is worth pointing out that, when (\ref{LaA}) and (\ref{LaB}) hold, the azimuthal component of the magnetic field also vanishes -- and not only the radial one --. This remark is important because previous analysis of the Meissner effect were based on the observation that the flux of the magnetic field through a hemisphere of the horizon vanishes, which of course suffices to verify the expulsion of the magnetic lines from the black hole. However, the vanishing of the azimuthal component in the near horizon limit can only be observed by explicitly computing the components of the field and not by comouting the flux. Figure \ref{fig:1} shows the lines of magnetic field and its strength close to the event horizon.  

\subsection{Thermodynamics of the Meissner state}

The neutral Meissner state we are interested in is characterized by (\ref{LaA}) and (\ref{LaB}), which in terms of the black hole charges corresponds to
\begin{equation}
e=0 \, , \ \ \ M^2=|j|\, .\label{urol}
\end{equation}
In such extremal uncharged state, the Smarr formula reduces to $M={2\Omega j}$, which, using the explicit expression for the angular velocity at the horizon, yields a remarkably simple expression for the product of the squared black hole mass and its entropy, namely 
\begin{equation}
M^2S={2\pi j^2}.
\end{equation}
Therefore, for the Meissner state we get that the mass and the entropy are
\begin{equation}
M=\sqrt{|j|} \, ,  \ \ \ \ \ S={2\pi |j|}\, .
\end{equation}
As we will see, this entropy can be reproduced with our near horizon approach.

\subsection{Charged black holes and Meissner effect}

\begin{figure}
\centering
\includegraphics[scale=0.50]{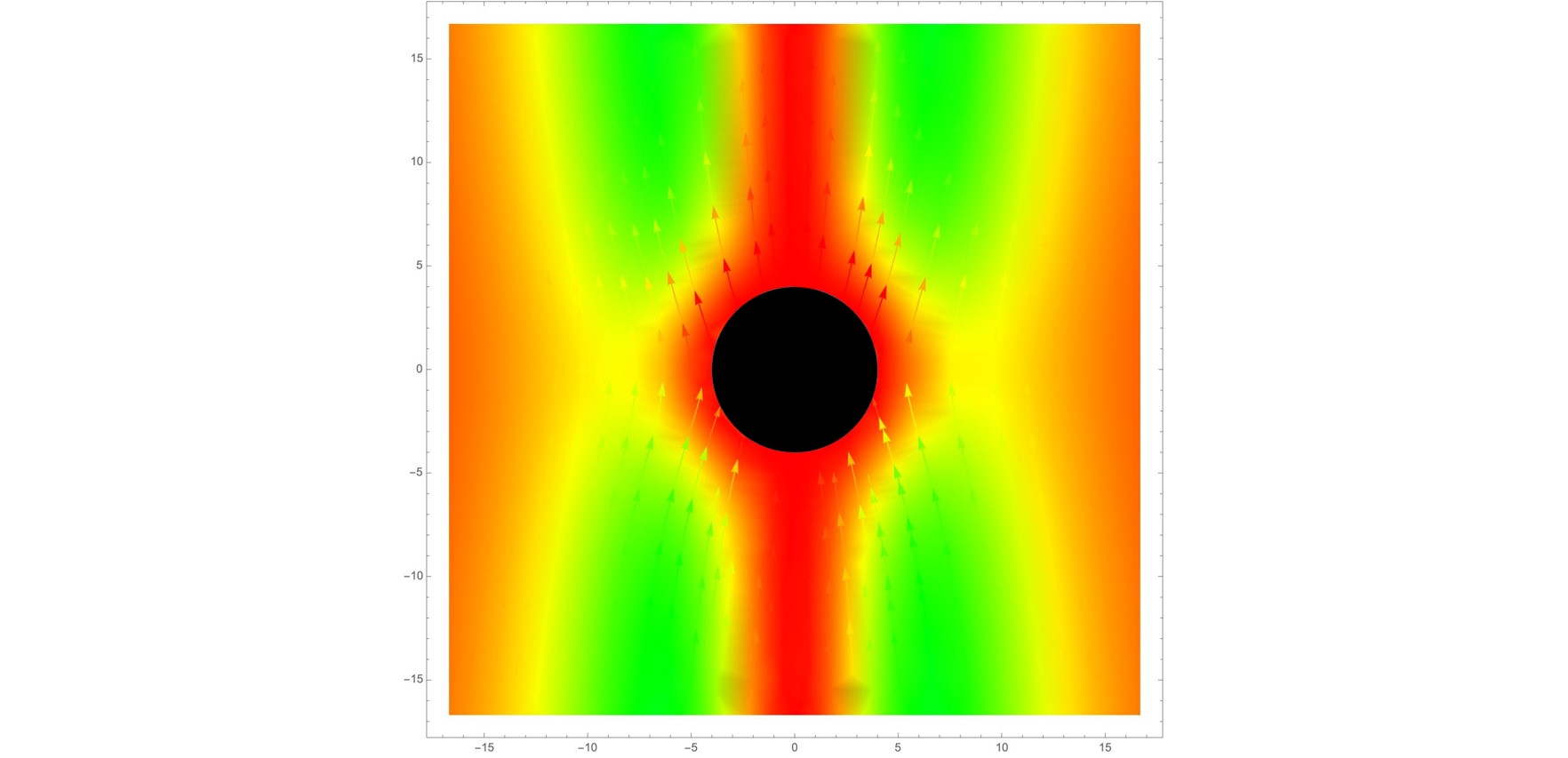}
\caption{Lines of magnetic field close to the event horizon for the case of an electrically charged black hole exhibiting Meissner effect. The colour code is the same as in the previous figures: red colour indicates the zones where the magnetic field is weaker, while yellow colour indicates where the magnetic field is stronger. The parameters of the black hole are such that $j=0$ and $M=|e|$.
The plots represent a transversal section of the solution, with the equatorial plane being perpendicular to the vertical axis. As the previous plot, this is in Cartesian coordinates, having adimensionalized.} \label{fig:2}
\end{figure}

Non-rotating black holes ($j=0$) in the Melvin background are also known to exhibit the Meissner effect \cite{Mac}, and we do observe it, as depicted in Figure \ref{fig:2}. 

While here we are mainly interested in the neutral extremal black holes, let us write down the charges in the electrically charged case for completeness: In fact, using the explicit expression for the electric potential at the horizon \cite{Astorino}, we obtain remarkably simple formulae for the mass and the entropy of the magnetized extremal non-spinning black hole ($j=0$); namely
\begin{equation}
M=|e| \, ,  \ \ \ \ \ S={\pi e^2}\, .
\end{equation}

\subsection{Symmetries and charges of the Meissner state}

In Section 2, we have shown that the Ernst-Wild solution to Einstein-Maxwell equations, describing a spinning black hole immersed in a magnetic Melvin universe, can be accommodated in a coordinate system that fulfill the near horizon asymptotic conditions studied in \cite{Puhm}. While we showed this for generic values of $m, a$ and $B$, the computation of the Noether charges (\ref{LACARGASSS}) was done for the non-extremal case ($\kappa \neq 0$), the extremal case being somehow special. In order to generalize the horizon charge computation to the extremal case, we revisit the results of \cite{Extended, Puhm}: On the constant-$v$ slices of the horizon, we consider a conformal metric
\begin{equation}
g_{AB}^{(0)} = \Theta\, \gamma_{AB}\, ,
\end{equation}
where $\gamma_{AB}$ is the metric of constant curvature on the 2-sphere and $A,B=1,2$ either refer to the coordinates $z^A=\{\varphi , \, \theta \}$ or to complex coordinates $z^A=\{z,\, \bar{z}\}$. The conformal factor $\Theta$ is given by an arbitrary function of these coordinates, say of $z$ and $\bar{z}$. Explicitly, we have
\begin{equation}
g_{AB}^{(0)} = \frac{2\Theta (z,\bar{z} )}{(1+|z|^2)^2}(\delta^{z}_{A}\delta^{\bar{z}}_{B} +\delta^{\bar{z}}_{A}\delta^{z}_{B})\, .
\end{equation}
In the extremal case ($\kappa =0$) the boundary conditions at the horizon are preserved by the asymptotic Killing vectors of the form 
\begin{equation}
\xi = D(z,\bar z )\, v\partial_v + Y(z)\partial_z + \bar Y (\bar z )\partial_{\bar z } +\mathcal{O}(\rho )\label{blaudiff}
\end{equation}
and by the gauge parameter 
\begin{equation}
\epsilon = U(z,\bar z ) -X(z,\bar z )\, vA^{(0)}_v +\mathcal{O}(\rho )\, ,\label{blaudifff}
\end{equation}
with $D(z,\bar z )$, $U(z,\bar z )$, $Y^A=\{Y(z), \bar{Y}(\bar z )\}$, $A=1,2$, being four arbitrary functions. Notice that $D=const$ corresponds to dilations of the null direction $v$; in the non-extremal case this gets replaced by local translations in $v$; this is the reason why diffeomorphisms generated by $D(z,\bar{z}) \, v\, \partial_v$ have non-vanishing commutator with the horizon supertranslations generated by $\chi = T(z,\bar z )\, \partial_v$, cf. \cite{Extended, Puhm}; schematically, the structure $[T,D]=D$ corresponds to translations and dilations on $H$. Vectors (\ref{blaudiff})-(\ref{blaudifff}) generate the change in the fields 
\begin{equation}
\delta_{\xi , \epsilon} g_{\mu \nu }=\mathcal{L}_{\xi }g_{\mu \nu}\, , \ \ \ \delta_{\xi , \epsilon} A_{\mu }=\mathcal{L}_{\xi }A_{\mu }+\partial_{\mu }\epsilon \, ,
\end{equation}
preserving the correct boundary conditions at the horizon ($\rho = 0$). The conserved charges associated to these symmetries are given by \cite{Extended}
\begin{equation}
Q[D, Y^{A}, U]= - \frac{1}{16\pi G }\int dS\, \Big( -2D+Y^{B}(g_{vB}^{(1)}+4A_{B}^{(0)}A_v^{(1)})+4UA^{(1)}_v  \Big)\, .\label{Laquesigue}
\end{equation}
While the charges $Q[D(z,\bar{z}),0,0]$ and $Q[0, 0,U(z,\bar{z})]$ generate two commuting copies of the level-0 current algebra $\hat{u}(1)_0$, i.e. supertranslations with no central extension, the charges $Q[0, \delta^A_{z}Y(z),0]$ and $Q[0,\delta^A_{\bar{z}}\bar{Y}(\bar{z}), 0]$ generate two copies of the 2-dimensional local conformal (Virasoro) algebra with vanishing central charge, i.e. two commuting copies of Witt algebra generated by $L(z)=Y^z(z)\partial_{z}$ and $\bar{L}(\bar{z})=Y^{\bar{z}}(\bar{z})\partial_{\bar{z}}$, with the Fourier expansion $L(z)=\sum_{n\in \mathbb{Z}}L_nz^n\,\partial_{{z}}$, $\bar{L}(\bar{z})=\sum_{n\in \mathbb{Z}}\bar{L}_n\bar{z}^n\,\partial_{\bar{z}}$ expressing the extended $SL(2, \mathbb{R})$ structure $[L,L]=L$, along with the non-diagonal piece $[L,D]=D$. The transformations generated by diffeomorphisms and gauge transformations associated to functions $D(z,\bar{z})$ and $U(z,\bar{z})$ are infinite-dimensional Abelian ideals of the full algebra; see \cite{Puhm} for details. One can verify that the charge $Q[1,0,0]$ reproduces the entropy for the exremal magnetized black hole; more precisely, we find
\begin{equation}
Q[1,0,0] = \frac{1}{2\pi }\, \frac{A}{4}\, .\label{ultimate}
\end{equation}
which is the entropy multiplied by a factor $1/(2\pi )$. For the neutral black hole in the Meissner state, we find that (\ref{ultimate}) yields
\begin{equation}
Q[1,0,0] = |j|\, .
\end{equation}

Notice that the factor $1/(2\pi )$ in (\ref{ultimate}) is reminiscent of the one appearing in the Kerr/CFT computation of the entropy \cite{AstorinoKerrCFT, otroKerrCFT}; in Kerr/CFT that factor is interpreted as coming from the left moving temperature in the Frolov-Thorne vacuum \cite{KerrCFT}. Understanding the precise connection between Kerr/CFT and our near-horizon calculations would be very interesting. The relation with other scenarios involving magnetic fields and extremal black holes, such as the analysis of the force-free electrodynamics done in \cite{Lupsasca:2014pfa}, is also worthwhile studying.

\section{Conclusions}

In this paper we have explicitly shown that Kerr-Newman black holes immersed in an external magnetic field exhibit infinite-dimensional symmetries in the near horizon limit. To show this, we applied the method developed in \cite{Laura, Extended, Puhm} to the Ernst-Wild solution to Einstein-Maxwell equations, which describes a spinning, electrically charged black hole embedded in a magnetic Melvin universe. By carefully adapting the formalism of \cite{Puhm} to the case of magnetized black holes, we wrote the asymptotic near horizon expansion for the spacetime metric and the gauge field; we showed that it corresponds to the boundary conditions yielding supertranslation and superrotation asymptotic symmetries at the horizon. Then, we showed that the Noether charges associated to the zero modes of these symmetries reproduce the physical variables of the magnetized black hole and its thermodynamics. This represents a generalization of the results of \cite{Lucho} to arbitrary values of the black hole parameters. In addition, we elaborated on the horizon symmetry computation by proving that the Noether charge associated to the angular momentum computed at the horizon admits to be expressed as the sum of two Komar integrals, one corresponding to the geometry contribution and one to the gauge field contribution. While the latter vanishes in the asymptotically flat spacetime, it does contribute to the angular momentum when the black hole is embedded in the Melvin magnetic bundle. We also showed the validity of the Gauss phenomenon by expressing the horizon charge associated to the electric charge as a flux integral. Then, we focused on the case in which the spinning black hole is neutral and it approaches extremality: this is the case in which the event horizon exhibits the Meissner effect. As explained in \cite{Extended}, for such configuration the horizon charges change, although the theory still exhibits an infinite-dimensional symmetry. This symmetry is still a combination of local conformal transformations and two sets of supertranslations. The latter correspond to superdilations on the null coordinate on the horizon and to gauge transformations that preserve the gauge field configuration at the horizon. The computation of the charges associated to the zero-modes of these symmetries allowed us to perform the analysis of the thermodynamics of the event horizon in its Meissner state reproducing the results in the literature.

To conclude, let us mention that an interesting future direction of this line of research would be to understand the near horizon description of the magnetized horizons from the perspective of \cite{Donnay}. There, the authors show that the geometry of a black hole horizon can be described as a Carrollian geometry emerging from an ultra-relativistic limit in the near-horizon region. Extending the formalism of \cite{Donnay} to study the dynamics of the magnetic field from the near horizon perspective would be interesting.

\[\]

The authors thank Marco Astorino, Sasha Brenner, Gregory Gabadadze, Andrei Gruzinov, and Juan Laurnagaray for discussions. This work has been partially supported by grants PIP-(2017)-1109, PICT-(2019)-00303, PIP-(2022)-11220210100685CO, PIP-(2022)-11220210100225CO, PICT-(2021)-GRFTI-00644.

\end{document}